\documentclass{an}
\usepackage{graphicx}
\usepackage{times}
\usepackage{fancyhdr}
\begin{document}

\title{Flip-flop phenomenon: observations and theory}

\author{D. Elstner\inst{1} 
\and  H. Korhonen\inst{1}}
\institute{
Astrophysikalisches Institut Potsdam, An der Sternwarte 16, 
D-14482 Potsdam, Germany}

\date{Received $<$date$>$; 
accepted $<$date$>$;
published online $<$date$>$}

\abstract{
In many active stars the spots concentrate on two permanent active longitudes
which are 180\degr\ apart. In some of these stars the dominant part of the
spot activity changes the longitude every few years. This so-called flip-flop
phenomenon has up to now been reported in 11 stars, both single and binary
alike, and including also the Sun. To explain this phenomenon, a
non-axisymmetric dynamo mode, giving rise to two permanent active longitudes
at opposite stellar hemispheres, is needed together with an oscillating
axisymmetric magnetic field. Here we discuss the observed characteristics of
the flip-flop phenomenon and present a dynamo solution to explain them.
\keywords{stars: activity -- stars: magnetic fields -- stars: spots --
  methods: numerical}
}

\correspondence{delstner@aip.de}

\maketitle

\section{Introduction}

The global structure and behaviour of the stellar magnetic fields are
determined by different dynamo modes that have different symmetries and
stabilities (see e.g. Brandenburg et al.~\cite{bra}). In slowly rotating
stars, like the Sun, axisymmetric modes are excited. These modes do not show
any structure in the longitudinal distribution of the spots and they
oscillate in time. In more rapidly rotating stars the higher order
non-axisymmetric modes get excited (see e.g. Moss et al.~\cite{mossetal};
Tuominen, Berdyugina~\& Korpi~\cite{tuo}). The magnetic configuration in the
non-axisymmetric modes consists of two starspots that are 180\degr\ apart,
explaining the permanent active longitudes seen in many rapidly rotating
stars. These non-axisymmetric modes do not oscillate. For explaining the
flip-flop phenomenon, where we see both active longitudes and oscillations,
axisymmetric dynamo modes need to co-exist with the non-axisymmetric modes.  

In this paper we describe the observed characteristics of the flip-flop
phenomenon and present a model that can produce them.

\section{Observations of flip-flop phenomenon}

\begin{figure}
\begin{center}
\resizebox{6cm}{!}
{\includegraphics{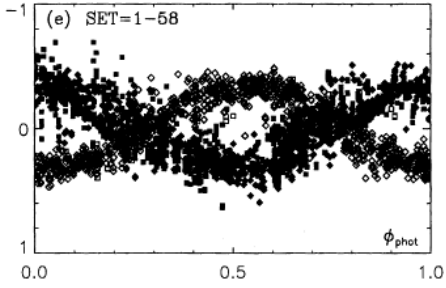}} 
\end{center}
\caption{The active longitude structure on FK~Com. Normalized magnitudes
  (ordinate) of FK Com for 1966--1990 plotted against the phase
  (abscissa). The phases have been determined using the ephemeris
  HJD~2439252.895 + 2.4002466~E. This figure has been taken from Jetsu et
  al.~(\cite{jetsu2}).}
\label{jetsu93}
\end{figure}

The flip-flop phenomenon, in which the main part of the spot activity changes
180\degr\ on the stellar surface, was first discovered in the early 1990s on a
single, very active, giant, FK Com. Jetsu et al.~(\cite{jetsu1},
\cite{jetsu2}) noticed from photometric observation, that the spot activity on
FK~Com for 1966--1990 concentrated on two longitudes, 180\degr\ apart. They
also noted that during the individual observing seasons only one of the active
longitudes had spots. This behaviour is very well illustrated in
Fig.~\ref{jetsu93} (from Jetsu et a;~\cite{jetsu2}), which shows the
normalized magnitudes of FK Com for 1966--1990. The photometric minimum is
always either around the phase 0.0 or 0.5. The data Jetsu et
al.~(\cite{jetsu2}) used can be analysed together with more recent
observations (1991--2003) to estimate the frequency at which flip-flop events
occur on FK~Com. There is on average one flip-flop event every 2.6 years,
giving full cycle length of 5.2 years (Korhonen et al.~\cite{korAN}). 

After the discovery of the flip-flop phenomenon on FK~Com it has been reported
also on other active stars. Berdyugina~\& Tuominen (\cite{ber_tuo}) studied the
photometric observations of four RS~CVn binaries and discovered that also
these stars have permanent active longitudes that are alternatively active. In
the case of II Peg Rodon{\`o} et al.\ (\cite{rod}) later confirmed their
results. Berdyugina, Pelt~\& Tuominen (\cite{ber_lqhya}) discovered
flip-flops on young solar type star, LQ~Hya. And a recent analysis of 120
years of sunspot data (Berdyugina \& Usoskin \cite{ber_sun}) suggests that the
Sun also has permanent active longitudes with associated flip-flops. On the
Sun a flip-flop event occurs on average every 3.8 years on the northern and
every 3.65 years on the southern hemisphere.

When the flip-flop phenomenon was discovered, it was not sure whether the
phenomenon was caused by spot movement across the stellar disk or emergence of
flux on the new active longitude. Korhonen et al.~(\cite{kor_ff}) have shown,
with Doppler images just before and after a flip-flop event on FK~Com, that
flip-flops are caused by changing the relative strengths of the spot groups at
the two active longitudes without actual spot movement on the stellar surface.

All the stars for which the flip-flop phenomenon has been reported are
listed in Table~\ref{fftable}. Spectral type, rotation period and the relative
differential rotation coefficient are given together with the flip-flop period
(the length of the full cycle). The flip-flop phenomenon has so far been
detected in many different kinds of stars: both binaries and single stars;
young, main sequence and evolved alike. Usually, the flip-flop period is
between 5 and 10 years, median being 7 years. The stars themselves usually have
rotation periods $<3$ days (median 2.4 days). Anyhow, no clear correlation
between the rotation period and the flip-flop period can be seen. 

Apart from the stars mentioned in Table~\ref{fftable}, there are also two other
stars for which flip-flops have been reported. These stars are single giant
HD199178 (Hackman \cite{hack}) and RS~CVn binary RT~Lac (Lanza et
al.~\cite{lanza}). In these stars only very few events have been observed, so
no information on the flip-flop cycle length can be obtained. 

\begin{table}[h]
\caption{Stars that show flip-flop phenomenon. In the Table the name of the
  star, spectral type and age, rotation period, flip-flop period and the
  relative differential rotation coefficient are given.}
\label{fftable}
\scriptsize{
\begin{tabular}{llccc}\hline
Name      &  Type               &$P_{\rm rot}$& $P_{\rm ff}$ & 
$\Delta\Omega/\Omega$ \\
\hline
Sun       & single, G2 V         & 27 d   & 7$^1$ yr    &  0.19 \\
LQ Hya    & single, K2 V, ZAMS   & 1.6 d  & 5.2$^2$ yr  &  0.022$^3$ \\
AB Dor    & single, K0 V, ZAMS   & 0.5 d  & 5.5$^4$ yr  &  0.05$^5$ \\
EK Dra    & single, G1.5 V, ZAMS & 2.6 d  & 4$^4$ yr  & - \\
FK Com    & single, G7 III       & 2.4 d  & 5.2$^6$ yr  &  0.018$^6$ \\
II Peg    & RS CVn, K2 IV        & 6.7 d  & 9.3$^7$ yr  &  0.04$^8$  \\
sigma Gem & RS CVn, K1 III       & 19.6 d & 14.9$^7$ yr & $<0.004^9$ \\
EI Eri    & RS CVn, G5 IV        & 1.95 d & 9.0$^7$ yr  & -0.15 --
-0.20$^{10}$ \\ 
HR 7275   & RS CVn, K1 III-IV    & 2.3 d  & 17.5$^7$ yr &   -  \\
\hline
\end{tabular}}
1)Berdyugina \& Usoskin \cite{ber_sun}
2)Berdyugina et al.~\cite{ber_lqhya}
3)K{\H o}v{\'a}ri et~al. \cite{kov2}
4)Berdyugina~\& J{\"a}r\-vi\-nen \cite{jarv}
5)Collier Cameron~\& Donati \cite{col}
6)Kor\-ho\-nen et al.~\cite{korAN} 
7)Berdyugina~\& Tuo\-mi\-nen \cite{ber_tuo}
8)Weber \cite{michi}
9)K{\H o}v{\'a}ri et~al. \cite{kov1}
10)Washuettl \cite{wasi}
\end{table}

\section{Modelling flip-flops}

The model consists of a turbulent fluid in a spherical shell of inner radius
$r_{\rm in}$ and outer radius $r_{\rm out}$.  

\begin{figure}
\begin{center}
\resizebox{4.5cm}{!}
{\includegraphics{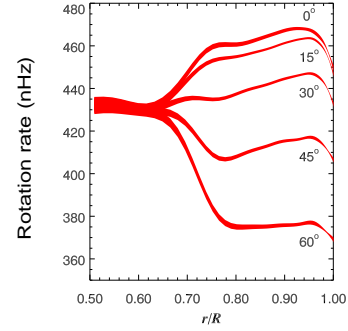}} 
\end{center}
\caption{The solar rotation law.}
\label{rot}
\end{figure}

We solve the induction equation 
\begin{equation}
{\partial \langle\vec{B}\rangle \over \partial t} = 
{\rm curl}(\alpha_q\circ \langle\vec{B}\rangle
-{\eta}_{\rm T} {\rm {curl}}\langle{\vec{B}}\rangle), 
\label{eq:1}
\end{equation} 
in spherical coordinates (${\rm r,\theta,\varphi}$) for an
$\alpha^2\Omega$-dynamo. A solar type rotation law (see Fig.~\ref{rot}) in the
corotating frame with the core 
\begin{eqnarray}
\lefteqn{\Omega(r,\theta)= 
{1\over 2}\Omega_0\left[1+{\rm erf}\left({r-r_{in} \over d_1}\right)\right]
(\Omega_{s} -\Omega_{c})}
\end{eqnarray} 
where $ \Omega_s=\Omega_{eq}- a {\rm cos}^2\theta $ is used. 

Only the symmetric part  
\begin{eqnarray}
\lefteqn{\alpha_{rr}=\alpha_0 {\rm cos}\theta(1.-2{\rm cos}^2\theta)}
\nonumber \\ 
\lefteqn{\alpha_{\theta\theta}=
\alpha_0 {\rm cos}\theta(1.-2{\rm sin}^2\theta)} \nonumber \\ 
\lefteqn{\alpha_{\varphi\varphi}=\alpha_0 {\rm cos}\theta} \nonumber \\
\lefteqn{\alpha_{r\theta}=\alpha_{\theta r}= 2 \alpha_0 {\rm
    cos}^2\theta{\rm sin}\theta}  
\end{eqnarray} 
of the $\alpha$-tensor is included. In order to saturate the dynamo we choose
a local quenching of 
\begin{equation}
\alpha_q={\alpha \over 1+\vec{B}^2/\vec{B}_{\rm eq}^2}$$
\label{eq:quench}
\end{equation}
For $\alpha_0$ we choose a value slightly above the critical one for the 
dynamo threshold. 

\begin{figure*}[ht]
\resizebox{4.99cm}{!}
    {\includegraphics*[bbllx=54pt,bblly=373pt,bburx=337pt,bbury=643pt]
      {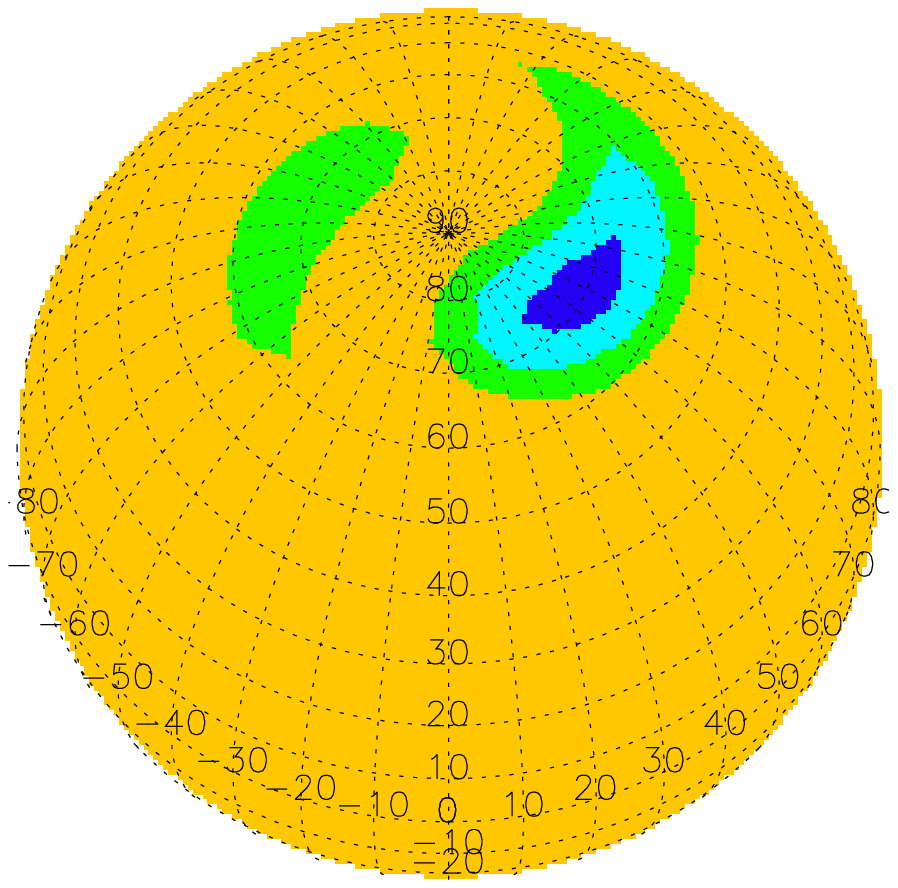}}
\resizebox{4.99cm}{!}
    {\includegraphics*[bbllx=54pt,bblly=373pt,bburx=337pt,bbury=643pt]
      {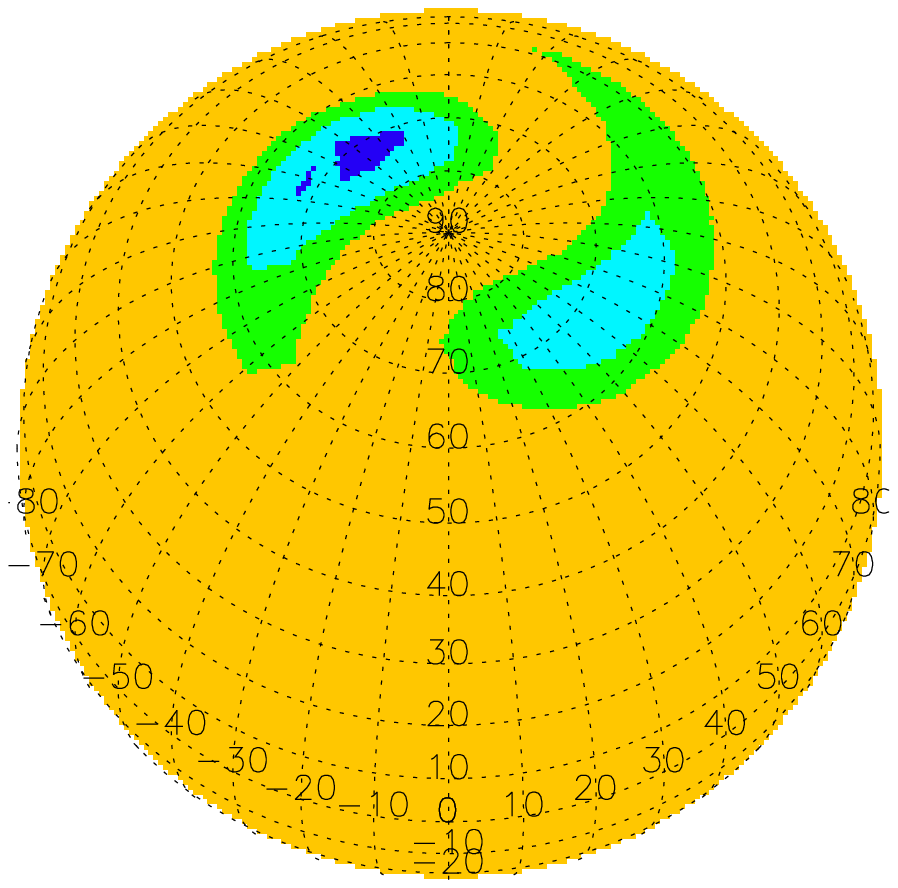}}
\resizebox{4.99cm}{!}
    {\includegraphics*[bbllx=54pt,bblly=373pt,bburx=337pt,bbury=643pt]
      {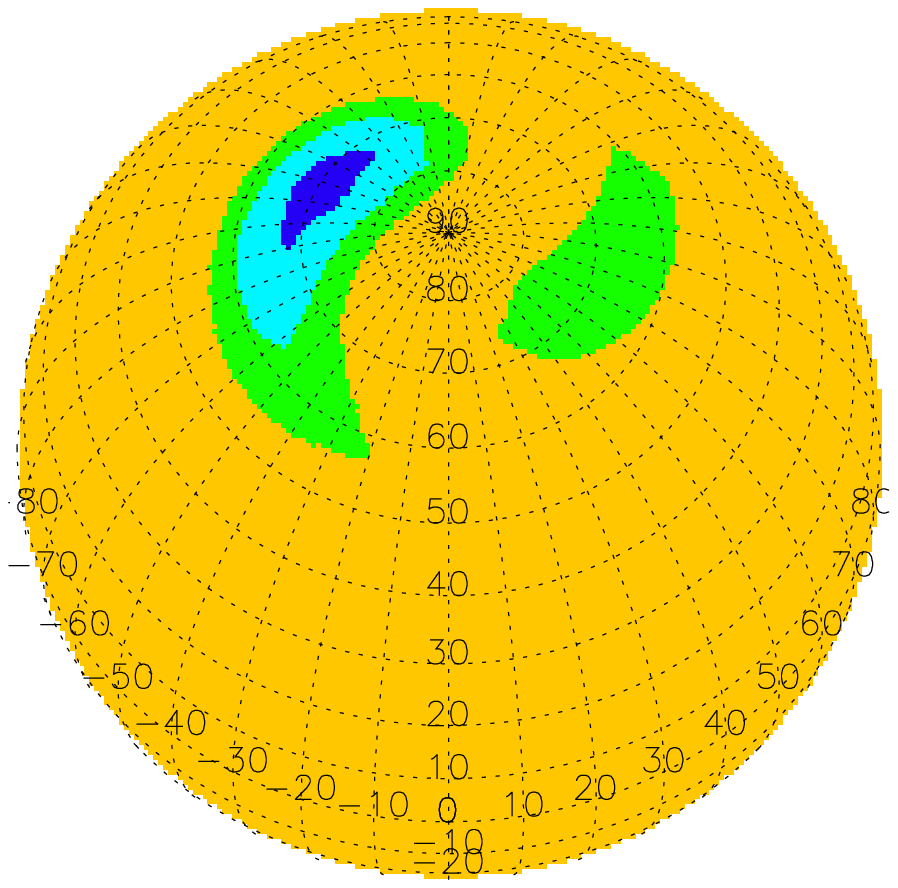}}    
\caption{The magnetic pressure from the dominant radial component 
of the magnetic field on the stellar surface is shown at three different time
steps for the thin layer model. We subtracted rotation and migration.} 
\label{thin}
\end{figure*}

\begin{figure*}[ht]
\resizebox{4.99cm}{!}
    {\includegraphics*[bbllx=54pt,bblly=373pt,bburx=337pt,bbury=643pt]
      {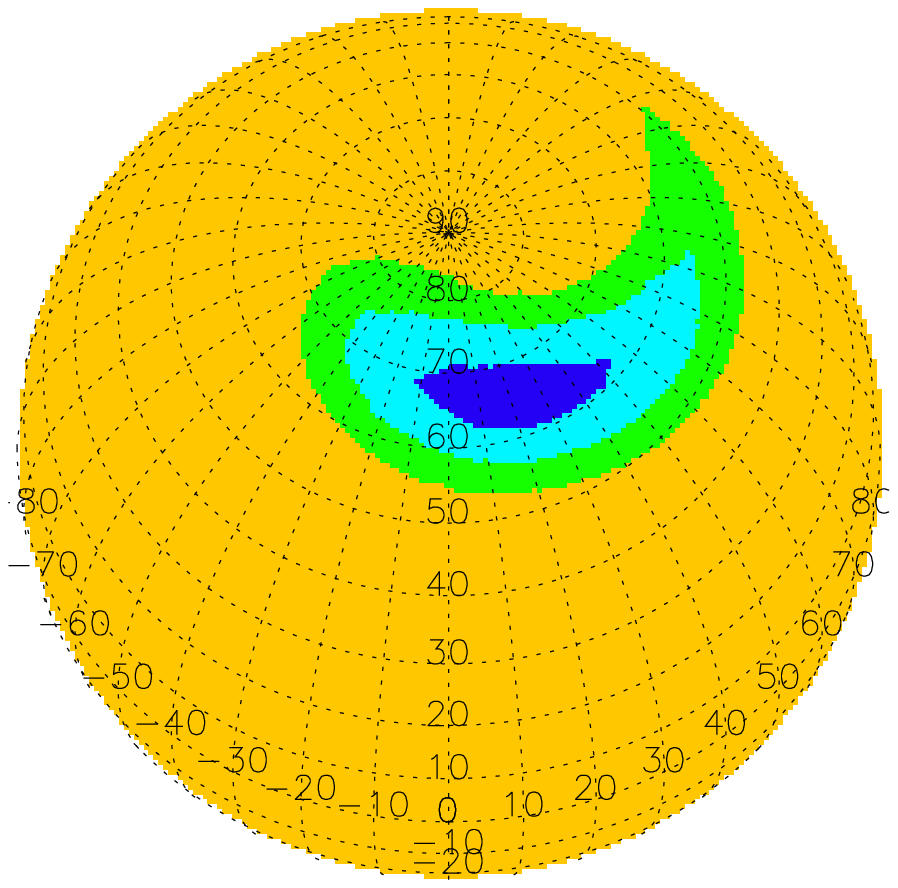}} 
\resizebox{4.99cm}{!}
    {\includegraphics*[bbllx=54pt,bblly=373pt,bburx=337pt,bbury=643pt]
      {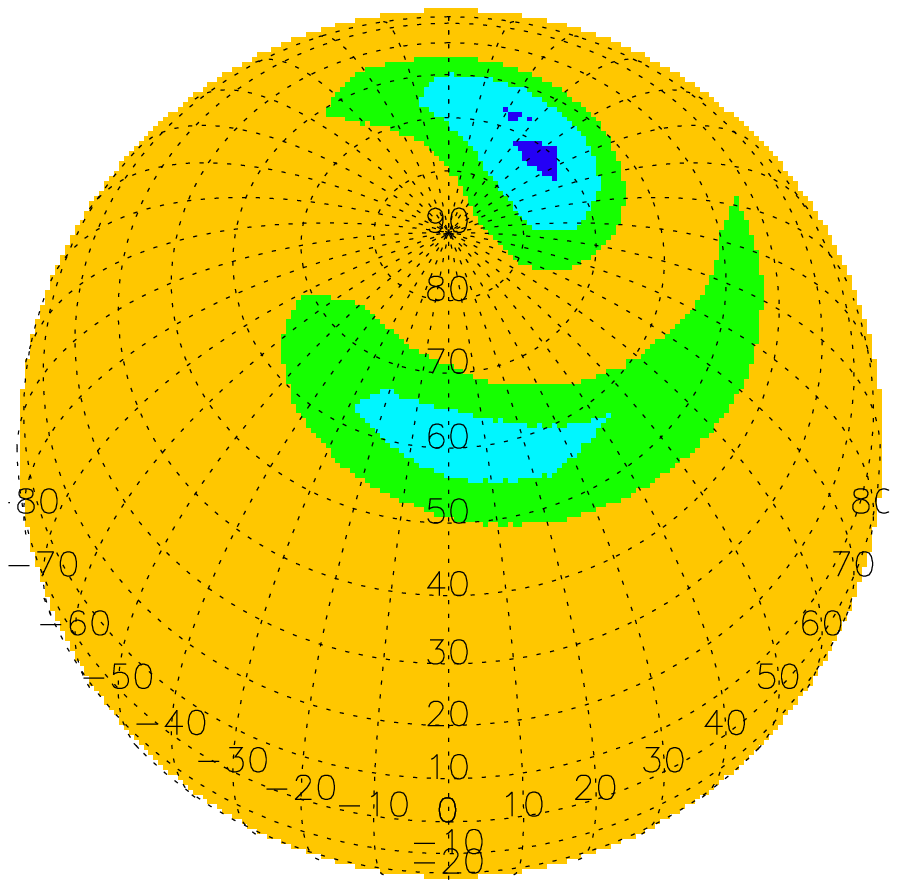}}
\resizebox{4.99cm}{!}
    {\includegraphics*[bbllx=54pt,bblly=373pt,bburx=337pt,bbury=643pt]
      {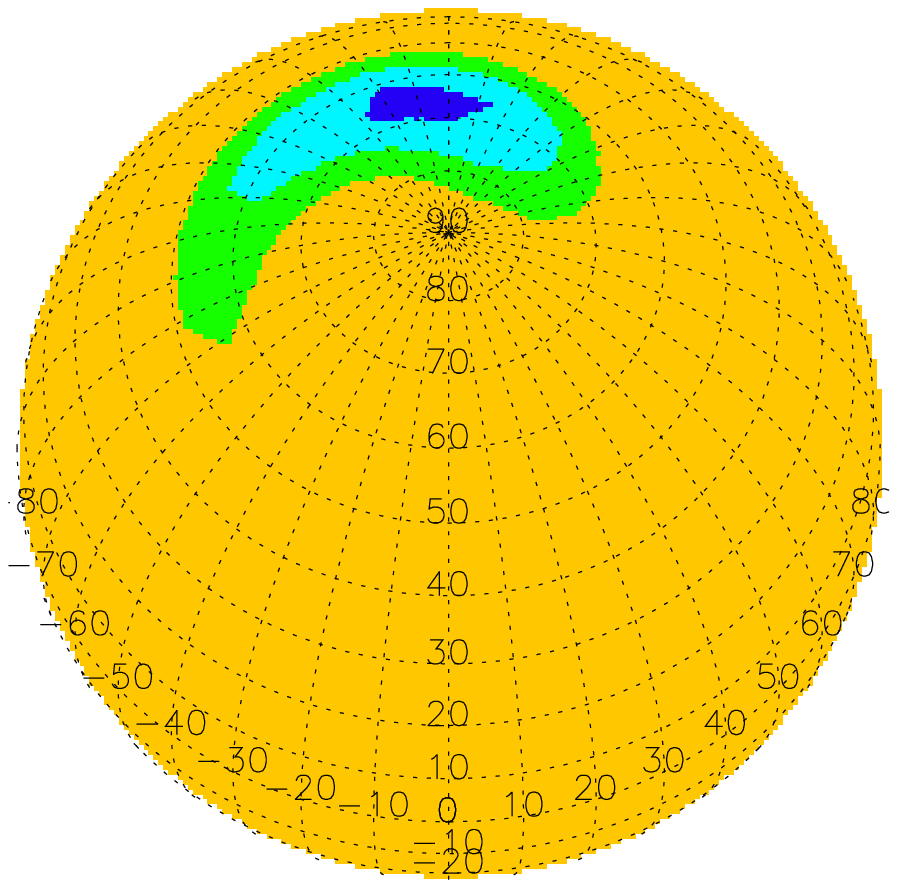}}\\
\resizebox{4.99cm}{!}
    {\includegraphics{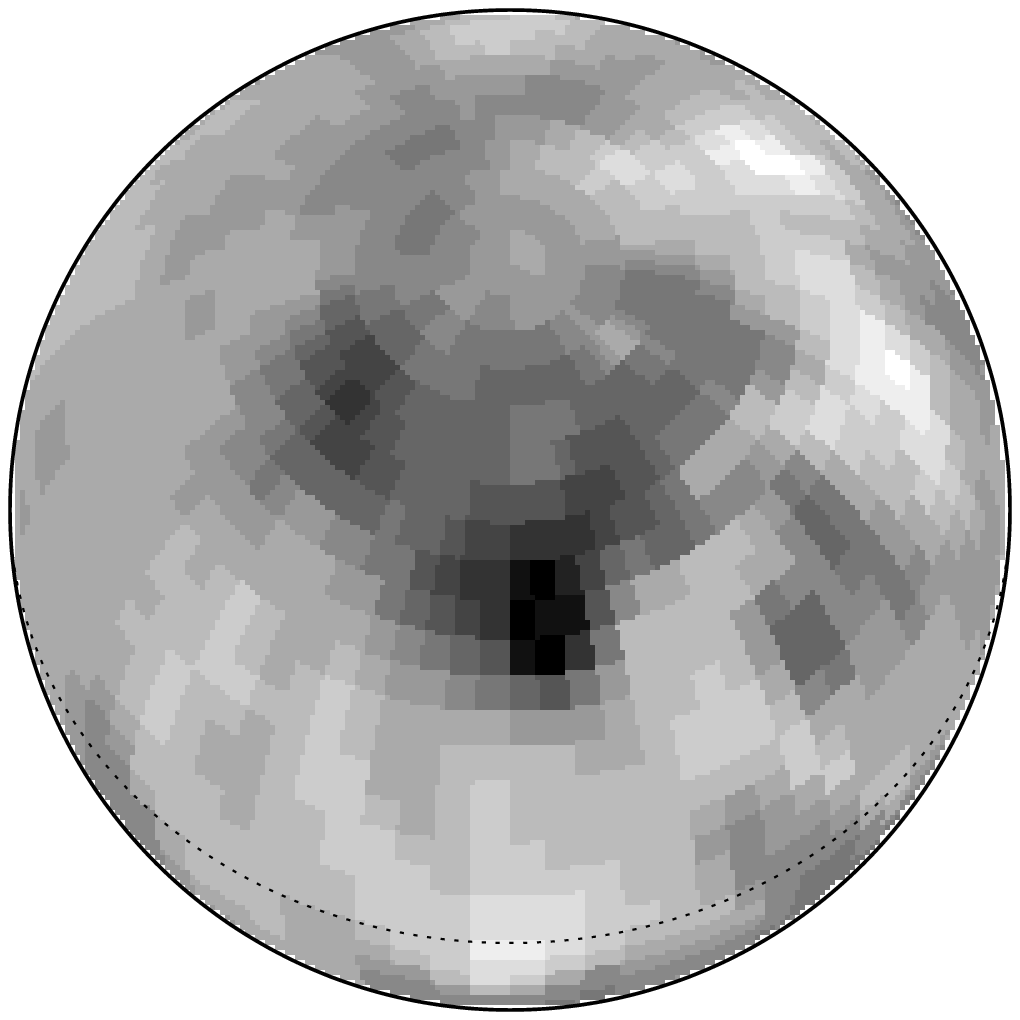}}
\resizebox{4.99cm}{!}
    {\includegraphics{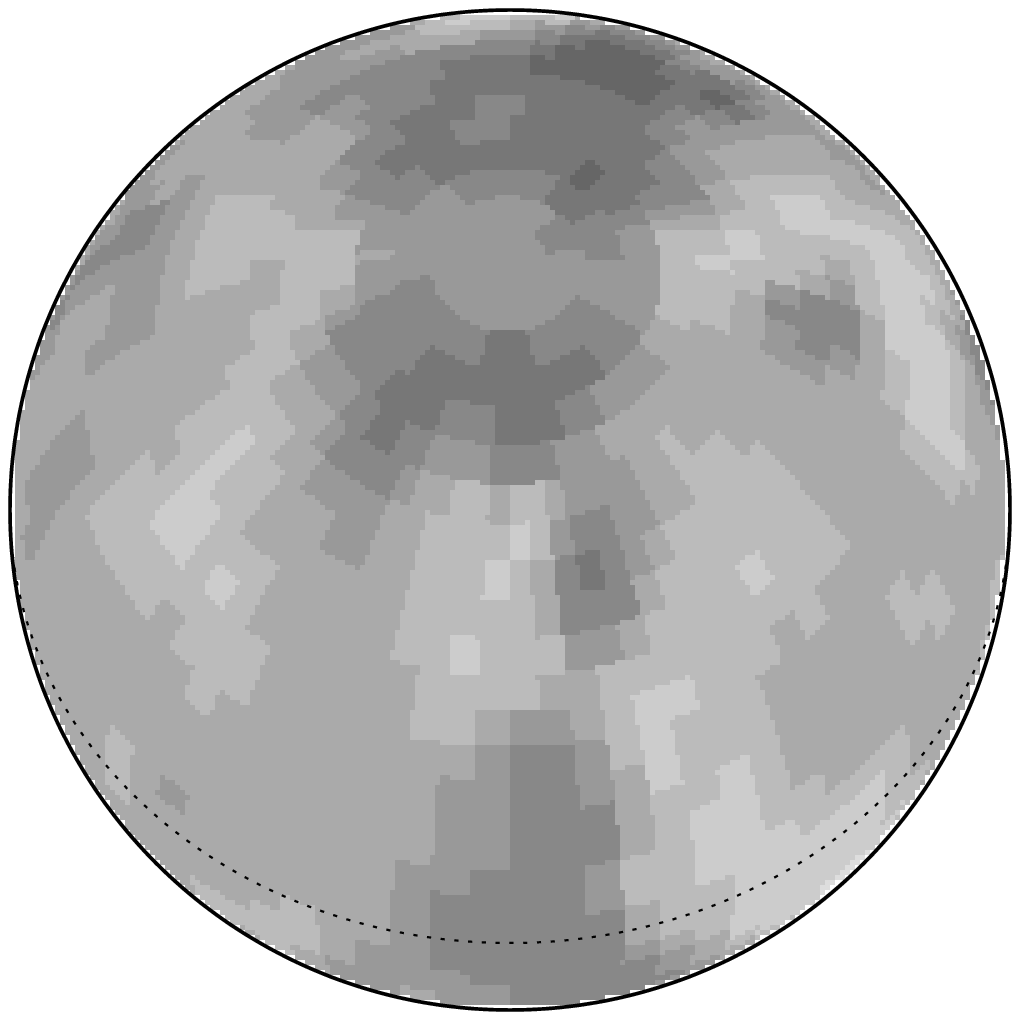}}
\caption{
The upper panel is as in Fig.~\ref{thin} but now for the thick
 convection zone model. The lower panel shows Doppler images of FK Com for
 June 1997 and January 1998 (from Korhonen et al.\ 2001). For the Doppler
 images the grey scale gives the temperature scale of 3600 K -- 5700 K. }
\label{thick}
\end{figure*}

The inner boundary is a perfect conductor and the outer boundary 
resembles a vacuum condition, by including an outer region up to 1.2 stellar
radii into the computational grid with 10 times higher diffusivity. At the
very outer part the pseudo vacuum condition is used. In order to see the
influence of the thickness of the convection zone we have chosen $r_{\rm
  in}=0.7$ for a thin (results shown in Fig.~\ref{thin}) and $r_{\rm in}=0.4$
for a thick (Fig.~\ref{thick}) convection zone. 

\section{Results}

With the parameter $\Omega_0$ we model the strength of the differential 
rotation.  For $\Omega_0=1$ we have a solar rotation law with an 
oscillating axisymmetric dynamo solution. The case $\Omega_0=0 $ is 
an $\alpha^2$-dynamo which gives a migrating non-axisymmetric dynamo 
because of the anisotropic $\alpha$ (cf. R{\"u}diger, Elstner~\&
Ossendrijver~\cite{rued}).

For 10\% of the solar differential rotation we found similar excitation
conditions for a drifting non-axisymmetric mode and an oscillating
axisymmetric mode. Because of the chosen positive $\alpha$ in the northern
hemisphere, we get a poleward migration of the oscillating mode. The drift of
the non-axisymmetric mode is opposite to the rotation. 

\begin{figure}[ht]
\resizebox{7.49cm}{!}
    {\includegraphics{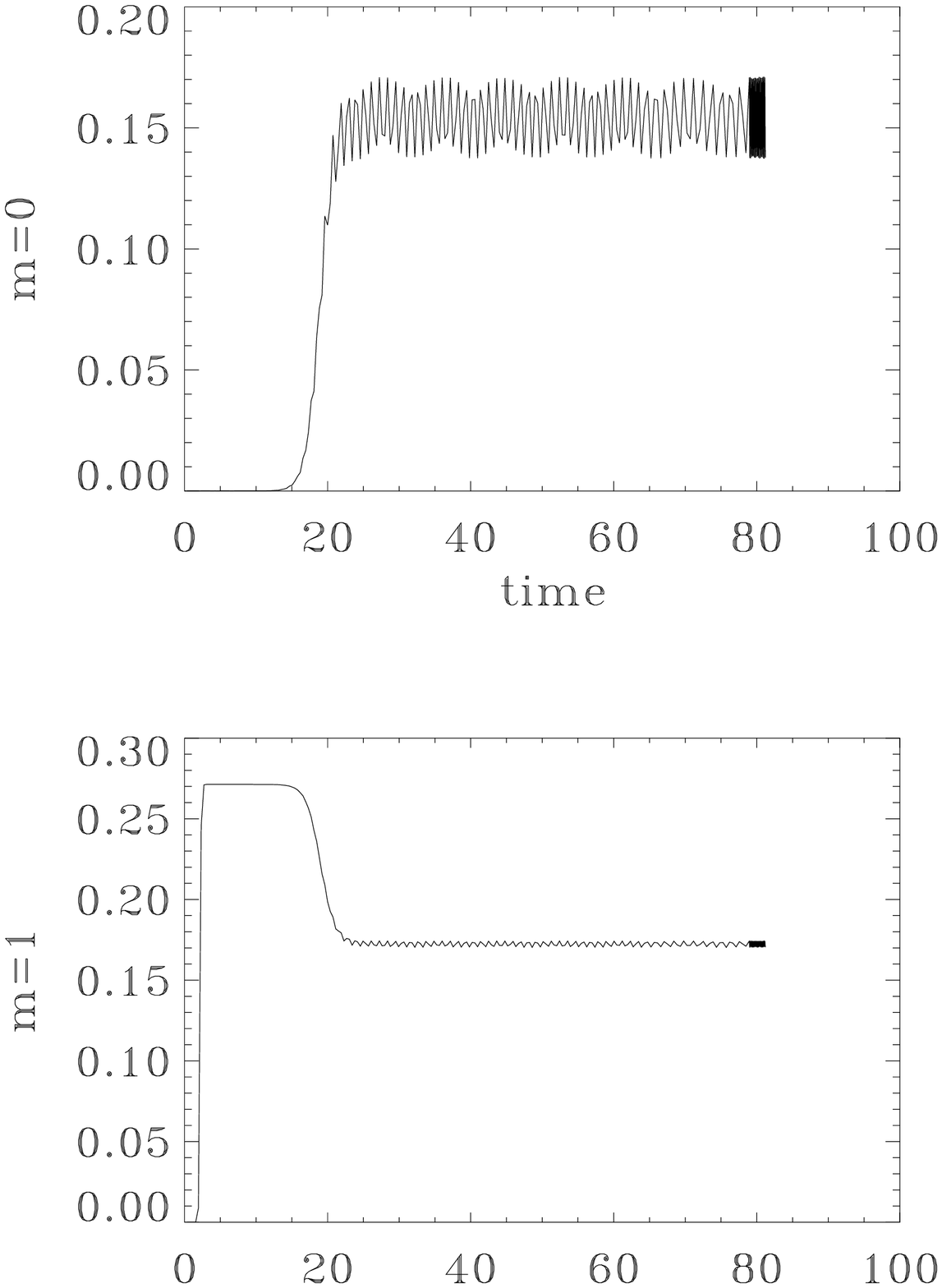}
    \includegraphics{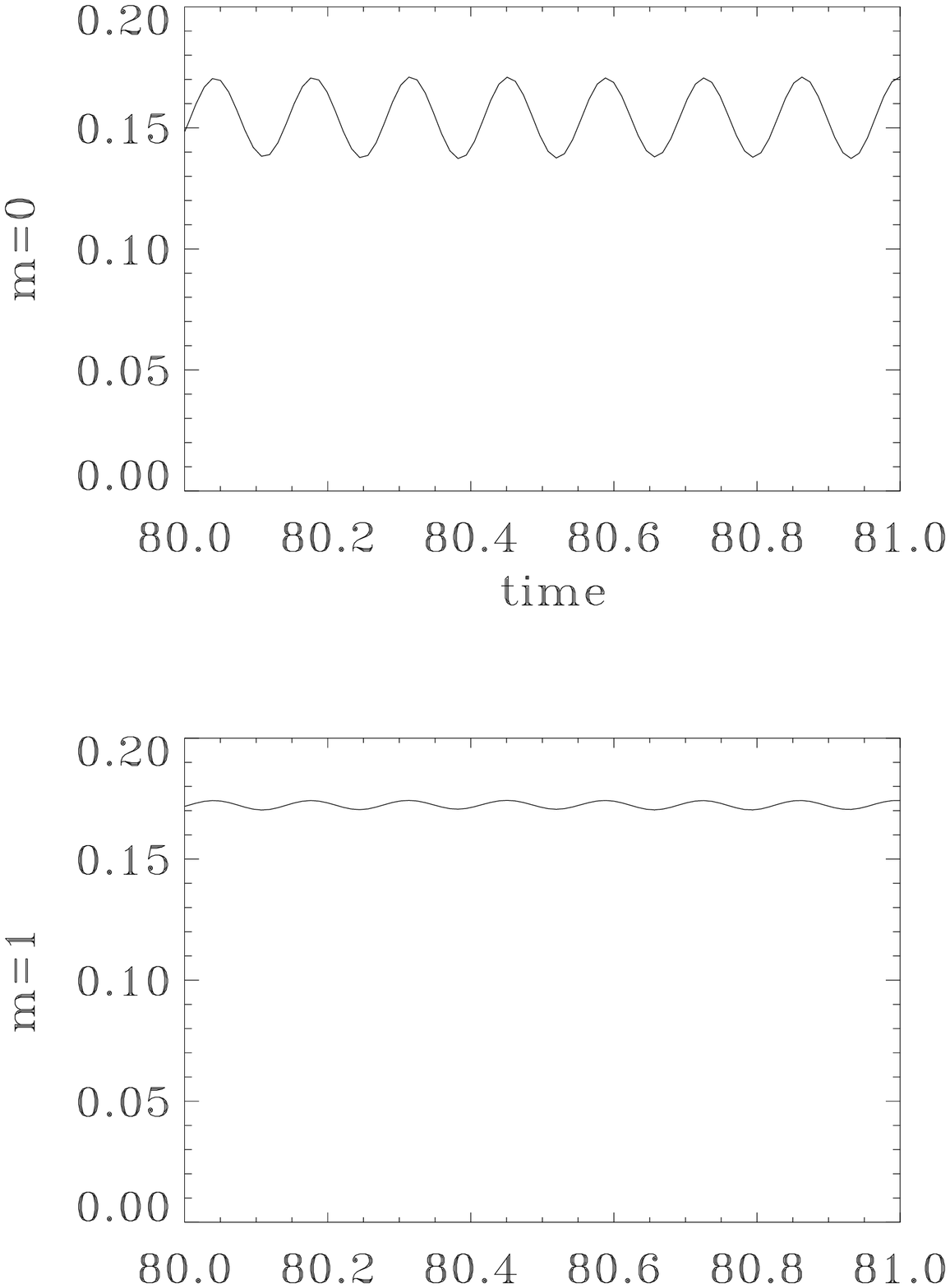}}
\caption{Energy density  in $m=0$ and $m=1$ modes for the thick model 
         normalized to equipartition. The time unit is the diffusion time of
         30 years. Left: The whole simulation Right: Final time with higher
         resolution. Notice the weak oscillation of the energy in the $m=1$
         mode synchronized with $m=0$ due to  $\alpha$-quenching. The
         migration period in Table~\ref{param} is directly taken from the $m=1$
         field.}
\label{figen}
\end{figure}

Using a simple $\alpha$-quenching, given by Eq.~\ref{eq:quench}, in 3D
simulations we found coexisting solutions for both modes, showing a magnetic
flip-flop phenomenon. We followed the solution in our simulation up to 100
diffusion times. There were no sign for it being only a temporary
phenomenon. The temporal behaviour of the magnetic energy is shown in
Fig.~\ref{figen}.
 
For an assumed turbulent diffusivity of about $10^{12}{\rm cm}^2{\rm s}^{-1}$
we get a period of about 6 years for the thin and 9 years for the thick model. 
These values are mainly determined by the magnetic diffusivity and vary only
weakly with the value for $\alpha$. The field strength saturates about the
equipartition value. In Table 2 we present a choice of models in order to
illustrate the parameter dependence of different solutions for an axisymmetric
(first row), a non-axisymmetric (second row) and two flip-flop (thick and thin
model; third and fourth row, respectively) solutions. Notice, that the
axisymmetric solution appears already for $\Omega_0=0.1$ but with a higher 
$\alpha$ than is used for the thick convection zone flip-flop model (third
row). This is probably due to the local quenching of the $m=1$ mode. 

\begin{table}
\caption{Parameters for a sample of runs with the energy density in 
equipartition units E0 for the mode $m=0$ and E1 for $m=1$, respectively. 
P0 denotes the period of the oscillation in diffusion times (30 years) and P1
the migration period. The dynamo-number $C_{\alpha}={\alpha_0 r_{star}
  \eta^{-1}}$ is used. }
\begin{tabular}{llccccc}\hline
$r_{\rm in}$ & $\Omega_0$ & $C_{\alpha}$ &   E0 & E1 & P0 & P1 \\ 
\hline
0.4    &      0.1   &    20      &   1.4    & $10^{-9 }$ &  0.27  &       \\
0.4    &      0.11  &    10      &  $10^{-6}$ &  0.1     &        & 0.23   \\
0.4    &      0.12  &    10      &   0.06   &  0.06    &  0.3   & 0.2    \\
0.7    &      0.11  &    21      &   0.1    &  0.4     &  0.23  & 0.46    \\
\hline
\end{tabular}
\label{param}
\end{table}

For the thick convection zone we found models where the magnetic spots appear
already at $50^{\circ}$ latitude and are migrating polewards during the
cycle. In this case the opposite spot does not start to appear exactly
$180^{\circ}$ away from the old spot. The distance can shrink down to
$90^{\circ}$. It depends also on $\alpha$. For weakly over critical $\alpha$
we find nearly $180^{\circ}$ distance between old and new spot. In all models
we found a counter-rotating migration of the magnetic pattern. The period was 
twice the flip-flop oscillation period for the thin layer and nearly equal to
the flip-flop period for our thick layer model. This migration should be
carefully considered in the future data analysis. 

\subsection{Comparison to FK Com}

Korhonen et al. (\cite{korAN}) found that for FK Com the relative surface
differential rotation ($\Delta\Omega/\Omega$) is about 10\% of the solar
value. This implies that the 10\% scaling of the solar rotation law, that has
been used in the model, is an appropriate first approximation for this
star. In Fig.~\ref{thick} the modeling results for the thick convection zone
model are compared to the observed surface temperature distributions on FK~Com
for June 1997 and January 1998 (Korhonen et al.~\cite{kor_ff}). In the
calculations the spots appear at high latitudes. This is also consistent with
the Doppler imaging results for FK~Com, which show spots mainly at high
latitudes, and basically never lower than latitude $45^{\circ}$. Also, the
spot size looks reasonable and is in line with the observations. 

\section{Discussion}

For the first time we found stable mixed mode solutions for weakly 
differential rotating stars. We could follow the cycle over 100 diffusion 
times. 

Moss (\cite{moss}) presented a similar model with isotropic $\alpha$. Because
there is no preferred m=1 mode in that case, it is probably not possible to
have stable mixed mode solutions for a longer time.  Also he did not find
similar values for the magnetic energy in both modes. In contrary to our
solution he got an asymmetric distribution of m=1 and $m=0$ modes  with
respect to the equator. We observed a similar exotic behaviour for highly
over-critical $\alpha$. 

The assumptions for the model are somewhat uncertain. First, changing the
diffusivity would change the period of the oscillating mode and therefore also
the flip-flop period. The simple scaling of the solar rotation law may not be
the best approximation for the stars showing flip-flops, also the simple
$\alpha$-quenching non-linearity may not be adequate. Anyhow, the main
properties remain for different quenching forms. Also, the simple diagonal
form of the $\alpha$ tensor is not justified. Nevertheless, the results from
our model are encouraging.

The flip-flop phenomenon appears in a limited range of the strength of 
differential rotation. The thickness of the convection zone is not very 
important. To what extent a large scale meridional flow changes this behaviour
has to be investigated. 

\acknowledgements
This project has been supported by the Deutsche Forschungsgemeinschaft grant
KO 2320/1. This research has made use of the Simbad database, operated at the
CDS, Strasbourg, France.


\begin{thebibliography}{}
\bibitem[1998]{ber_tuo} 
Berdyugina, S.V., Tuominen, I.: 
1998, A\&A 336, L25
\bibitem[2002]{ber_lqhya} 
Berdyugina, S.V., Pelt, J., Tuominen, I.: 
2002, A\&A 394, 505
\bibitem[2003]{ber_sun} 
Berdyugina, S.V., Usoskin, I.: 
2003, A\&A 405, 1121
\bibitem[2005]{jarv} 
Berdyugina, S.V., J{\"a}rvinen, S.P.: 
2005, this volume of AN
\bibitem[1989]{bra} 
Brandenburg, A., Krause, F., Meinel, R., Moss, D., Tuominen, I.:
1989, A\&A 213, 411
\bibitem[2002]{col} 
Collier Cameron, A., Donati, J.-F.: 
2002, MNRAS 329, L23
\bibitem[2004]{hack}
Hackman, T.: 
2004, A\&A submitted
\bibitem[1991]{jetsu1} 
Jetsu, L., Pelt, J., Tuominen, I., Nations, H.L.: 1991,
in: The Sun and Cool Stars: activity, magnetism, dynamos, 
Tuominen I., Moss D., R\"udiger G.\ (eds.), Proc.\ IAU Coll.\ 130,
Springer, Heidelberg, p.\ 381
\bibitem[1993]{jetsu2} 
Jetsu, L., Pelt, J., Tuominen, I.: 
1993, A\&A 278, 449
\bibitem[2001]{kor_ff}
Korhonen, H., Berdyugina, S.V., Strassmeier, K.G., Tuominen, I.: 
2001, A\&A, 379, L30
\bibitem[2004]{korAN}
Korhonen, H., Berdyugina, S.V., Tuominen, I.: 
2004, AN 325, 402
\bibitem[2001]{kov1} 
K{\H o}v{\'a}ri, Zs., Strassmeier, K. G., Bartus, J., Washuettl, A., Weber,
M., Rice, J.B.:  
2001,  A\&A 373, 199  
\bibitem[2004]{kov2} 
K{\H o}v{\'a}ri, Zs., Strassmeier, K. G., Granzer, T., Weber, M., Ol{\'a}h,
K., Rice, J.B: 
2004, A\&A 417, 1047 
\bibitem[2002]{lanza}
Lanza, A.F., Catalano, S., Rodon{\`o}, M., \.{I}bano\u{g}lu, C., Evren, S.,
Ta\c{s}, G., \c{C}ak{\i}rl{\i}, \"{O}., Devlen, A.: 
2002, A\&A 386,583
\bibitem[1995]{mossetal} 
Moss, D., Barker, D.M., Brandenburg, A., Tuominen, I.: 
1995, A\&A 294, 155
\bibitem[2004]{moss}
Moss, D.:
2004, MNRAS 
\bibitem[2000]{rod} 
Rodon{\`o}, M.,  Messina, S., Lanza, A.F., Cutispoto, G., Teriaca, L.: 
2000, A\&A 358, 624  
\bibitem[2003]{rued}
R{\"u}diger, G., Elstner, D., Ossendrijver, M.:
2003, A\&A 406, 15
\bibitem[2002]{tuo} 
Tuominen, I., Berdyugina, S.V., Korpi, M.J.: 
2002, AN 323, 367
\bibitem[2004]{wasi} 
Washuettl, A.:
2004, PhD Thesis, University of Potsdam
\bibitem[2004]{michi}
Weber, M.: 
2004, PhD Thesis, University of Potsdam
\end{thebibliography}
\end{document}